\documentclass[epj]{svjour}
\usepackage{graphics}
\usepackage{epsfig,amsmath,amssymb,float}   
\begin{document}
\input{epsf}
\newfloat{figure}{ht}{aux}
\newcommand{\bc}{\begin{center}}
\newcommand{\ec}{\end{center}}
\newcommand{\be}{\begin{equation}}
\newcommand{\ee}{\end{equation}}
\newcommand{\beqn}{\begin{eqnarray}}
\newcommand{\eeqn}{\end{eqnarray}}
\title{Scaling of the spin stiffness in random spin-$\frac{1}{2}$ chains}
\subtitle{Crossover from pure-metallic behaviour to random 
singlet-localized regime}

\author{Nicolas Laflorencie\inst{1} \and Heiko Rieger\inst{2}}

\institute{Laboratoire de Physique Th\'eorique, CNRS-UMR5152
Universit\'e Paul Sabatier, F-31062 Toulouse, France \and Theoretische Physik; Universit\"at des Saarlandes; 66041 Saarbr\"ucken; Germany}  

\date{\today}

\abstract{In this paper we study the localization transition induced 
by the disorder in random antiferromagnetic spin-$\frac{1}{2}$ chains.
The results of numerical large scale computations are presented for the
XX model using its free fermions representation. The scaling behavior
of the spin stiffness is investigated for various disorder strengths.
The disorder dependence of the localization length is studied and a
comparison between numerical results and bosonization arguments is
presented. A non trivial connection between localization effects and
the crossover from the pure XX fixed point to the infinite randomness
fixed point is pointed out.}
\maketitle

\section{Introduction}

Quantum spin chains exhibit a large number of interesting features
because the quantum fluctuations are often relevant, especially at low
temperature. The antiferromagnetic (AF) Heisenberg model in one
dimension (1D) has been extensively studied since the discovery in
1931 of the Bethe Ansatz \cite{Bethe31} for the spin $S=\frac{1}{2}$
chain. In 1D, the AF XXZ model defined by the Hamiltonian
\be
\label{eq:DefHamXXZ}
{\mathcal{H}}^{XXZ}=J\sum_{i=1}^{L}\Bigl[\frac{1}{2}(S_i^+S_{i+1}^-+{\mathrm h.c})+\Delta
S_i^zS_{i+1}^z\Bigr]
\ee
with $J>0$ and $\Delta \ge 0$, exhibits a gap-less excitation spectrum
for $S=\frac{1}{2}$ if $\Delta \le 1$, whereas a gap opens up in the
spectrum when $\Delta > 1$. In 1D, the quantum fluctuations prevent
the formation of true long-range order~\cite{Mermin66} but in the
critical regime $\Delta \le 1$ the model [Eq.(\ref{eq:DefHamXXZ})]
displays a {\it {quasi}}-long-range order (QLRO) with power-law
decaying spin-spin correlation functions in the ground state (GS). It
is well known that the model [Eq.(\ref{eq:DefHamXXZ})], without
quenched disorder, is integrable for conventional periodic boundary
conditions \cite{Bethe31} as well as in the more general
case of twisted boundary conditions (TBC) \cite{TBC}. The latter are
defined by:
\be
\label{eq:TBC}
S^z_{L+1}=S^z_1,\ \  S^{\pm}_{L+1} = S^{\pm}_1 e^{\pm i\phi},
\ee
where $\phi$ is the twist angle and is equivalent to a ring of
interacting fermions threaded by a magnetic flux of strength
$\frac{\hbar c}{e}\phi$ \cite{byers61}. The spin stiffness $\rho_S$ is
defined by
\be
\label{eq:stiff.def}
\rho_S=L^2 \frac{\partial^2 \epsilon_0(\phi)}{\partial\phi^2}|_{\phi=0},
\ee
where $\epsilon_0$ is the GS energy per site. It measures the
magnetization transport along the ring and in the fermionic language
this is called the charge stiffness, which is the Drude
weight of the conductivity. The gap-less phase is characterized by
peculiar transport properties: in the thermodynamic limit
Shastry and Sutherland \cite{shas90} showed that in the critical
regime the spin stiffness of the XXZ chain follows:
\be
\label{eq:stif.pure}
\rho_S(\Delta)=J\frac{\pi \sin(\mu)}{4 \mu (\pi -\mu)} \mathrm{~~where}\ \Delta=\cos(\mu),
\ee
and it vanishes for $\Delta > 1$.  The phase transition which occurs
at $\Delta=1$ can be viewed as a metal-insulator
transition~\cite{shas90} between a critical metallic phase with a
finite $\rho_S$ and a gaped insulating regime where $\rho_S=0$,
following a {\it{Mott}} mechanism.

When the system is not homogeneous, the situation described above
changes dramatically. For instance when only one coupling exchange
is weaker than the others in an otherwise homogeneous ring, the
stiffness has been found to scale to zero by numerical studies
\cite{weaklink02}, in perfect agreement with renormalization group
arguments of Eggert and Affleck \cite{EggAff92}, and Kane and Fisher
\cite{Kane92}. 

Moreover, for the case of a random spin-$\frac{1}{2}$ chain,
Doty and Fisher \cite{Doty92} performed a bosonization study
considering several types of random perturbations added to
(\ref{eq:DefHamXXZ}). They concluded that in the AF critical regime
the GS with QLRO is destroyed by any small amount of disorder and the phase
transition associated is an {\it{Anderson}} localization transition
\cite{anderson}, reminiscent of the localization
problem in 1D disordered metals studied by Giamarchi and Schulz
\cite{Giamarchi88}. A relevant length scale associated with the Anderson
transition is the localization length $\xi^*$. \\
More generally, the problem of transport in 1D random media
\cite{Huse} as well as localization effects and persistent currents in
disordered quantum rings have motivated a large number of theoretical
studies in recent years
\cite{Poilb94,Runge94,Giam95,Schmitteckert98,Urba03,Schuetz03}. In the
context of mesoscopic physics it turned out to be very interesting to
study the transport properties for finite systems, where coherence
effects are important~\cite{ian01,SCIENCE.NATURE}. In particular the
finite size (FS) dependence of the current, susceptibility and stiffness are
important for a complete understanding of the experimental results.

In the present paper we investigate the scaling behavior of the spin
stiffness of the random spin-$\frac{1}{2}$ chain. It is organized as
follows. In Sec.~2, the numerical method, based on the free fermions
formalism, is explained and notably the computation of the spin
stiffness is described. Sec.~3 is devoted to the study of the
localization transition: Using some bosonization arguments as well as
FS scaling analysis, an universal scaling of the stiffness to $0$ is
expected and we compare it with numerical results. In Sec.~4, the
disorder dependence of the localization length is studied and the
bosonization predictions are demonstrated to be valid only for weak
randomness. For strong disorder we propose a new quantity which gives
a better description for the disorder dependence of $\xi^*$. The
relation to crossover effects observed recently for spin-spin
correlation functions \cite{comment,laflo03} is also worked.
Sec.~5 contains some concluding remarks.

\section{Numerical method at the XX point}

We start with the 1D random XX model on a ring closed with TBC. It is
defined by
\begin{equation}
\label{eq:DefReXX}
{\mathcal{H}}_{\rm{random}}^{XX}(\phi)=\sum_{i=1}^{L-1}\Bigl[\frac{J_i}{2}(S_i^+S^{-}_{i+1}+{\mathrm {h.c}})\Bigr] +h_{L}(\phi),
\end{equation}
with the boundary term $h_{L}(\phi)=\frac{J_L}{2}(S^+_L S^-_1
e^{-i\phi} + {\mathrm {h.c}})$.  The couplings $J_i$ are independent
random numbers.

\subsection{Free fermions formulation}

For $S=\frac{1}{2}$ the well known Jordan-Wigner mapping transforms
spin operators into Fermi operators according to
\be
\label{eq:JW}
S^+_j = C_{j}^{\dagger} e^{i\pi\sum_{l=1}^{j-1}N_l},~~S^-_j = e^{-i\pi\sum_{l=1}^{j-1}N_l} C_{j}.
\ee
$N_j=C^{\dagger}_j C_j$ is the occupation number ($0$ or $1$) at site
$j$, given by $N_j=1/2+S^z_j$. Note that the Fermi anti-commutation
relations are satisfied $\{C^{\dagger}_i,C_j\}=\delta_{i,j}$.  The
Hamiltonian (\ref{eq:DefReXX}) can then be written as
\be
\label{eq:DefReXXFerm}
{\mathcal{H}}_{\rm{random}}^{XX}(\phi)= \sum_{i=1}^{L-1}\Bigl[\frac{J_i}{2}(C_{i}^{\dagger}C_{i+1}+C_{i+1}^{\dagger}C_{i})\Bigr]+h_{L}(\phi).
\ee
The sign of the boundary term depends on the parity of the total
number of fermions ${\mathcal N}=\sum_{i=1}^{L}N_j$; indeed
\be
\label{eq:hPhi}
h_{L}(\phi)=-e^{i\pi{\mathcal N}}\frac{J_L}{2}(C^{\dagger}_L C_1 e^{-i\phi} + C^{\dagger}_1 C_L e^{i\phi}).
\ee
Hence, when $\phi=0$ the resulting free fermions problem must have
anti-periodic boundary conditions if the number of fermions is even and
periodic boundary conditions if $\mathcal N$ is odd \cite{Young96}.  In the
non-random case, the solution of the problem via a Fourier
transformation is trivial~\cite{Lieb61} due to its translational
invariance. In $k$-space, the pure model is given by
\be
\label{eq:DefPureXX}
{\mathcal{H}}_{\rm{pure}}^{XX}=-J\sum_{k}C^{\dagger}_{k} C_k \cos(k).
\ee
Its GS is at half-filling ($\mathcal{N}=\frac{L}{2}$,
corresponding to the $S^{z}_{\mathrm{tot}}=0$ sector). The twist angle
at the boundary produces a shift in the momentum space $k\rightarrow
k+\phi$ which can be uniformly distributed over all bond resulting in
a local twist $\delta \phi = \frac{\phi}{L}$ for each bond. Therefore
the GS energy per site takes the following simple expression
\be
\label{eq:GsPureEnPhi}
\epsilon_{0}(L,\phi)=-\frac{J}{L} \sum_{p}\cos(\frac{2 \pi p}{L}+\frac{\phi}{L}) = -\frac{J}{L}\frac{\cos(\frac{\phi}{L})}{\sin(\frac{\pi}{L})}
\ee
from which we can easily extract the spin stiffness \cite{laflo01} :
\be
\label{FSSS}
\rho_{S}(L)=J(L\sin(\frac{\pi}{L}))^{-1}\simeq \pi^{-1}+{\cal
O}(L^{-2}).
\ee
When the system is inhomogeneous, the translational invariance is
broken and a solution in the reciprocal space is no longer
possible. Fortunately the problem can be easily diagonalized numerically, 
using standard linear algebra routines \cite{noteNum}.
Indeed, with an unitary transformation the hamiltonian (\ref{eq:DefReXXFerm}) 
can be expressed in a diagonal form \cite{Lieb61,Young96,Henelius98,Igloi00}.
For completeness we give here a
brief description of the method. First let us define a column vector
$\Psi$ of size $L$ and its conjugate row vector $\Psi^{\dagger}$ by
\begin{equation}
\label{eq:PsiDef}
\Psi^{\dagger}=(C_{1}^{\dagger},...,C_{L}^{\dagger}).
\end{equation}
Hence, using this notation, we can re-write the Hamiltonian
(\ref{eq:DefReXXFerm}) in terms of a symmetric $L\times L$ band matrix
${\mathcal A}(\phi)$ as
\begin{equation}
\label{matrix}
{\mathcal{H}}_{\rm{random}}^{XX} (\phi)= \Psi^{\dagger}~{\mathcal A}(\phi)~\Psi,
\end{equation}
with non-zero elements given by ${\mathcal A}_{i,i+1}=\frac{J_{i}}{2}$
and at the boundaries, ${\mathcal
A}_{1,L}=(-1)^{\mathcal{N}}\frac{J_{L}}{2}e^{-i\phi}$. One can define
the unitary transformation $P$ that diagonalizes ${\mathcal A}$. Then
we get a new set of Fermi operators $\eta_{q}$ defined by
\begin{equation}
\eta_q=\sum_i P_{iq}C_i,~~
\eta_q^{\dagger}=\sum_i {P^\dagger}_{iq}C_i^{\dagger},
\end{equation}
which yields the following diagonal form for the Hamiltonian 
\begin{equation}
{\mathcal{H}}_{\rm{random}}^{XX}(\phi)=\sum_{q=1}^L e_q (\phi)\eta_q^{\dagger} \eta_q,
\end{equation}
where the $e_q(\phi)$ are the eigenvalues of ${\mathcal A}(\phi)$. At
temperature $T$, the occupation number is given by the Fermi function
$\langle N_{q}\rangle=\langle\eta_q^{\dagger} \eta_q\rangle=(1+e^{e_{q}(\phi)/T})^{-1}$. Because of
the particle-hole symmetry, the eigenvalues occur in pairs, equal in
magnitude and opposite in sign. Hence, at $T=0$, the GS energy is
simply given by
\be
\label{eq:GSEn}
\epsilon_{0}(\phi)=\sum_{q=1}^{{\mathcal N}=L/2}e_q(\phi),
\ee
where $e_1(\phi)\le e_2(\phi)\le~...~\le e_{L}(\phi)$.

\subsection{Numerical evaluation of the spin stiffness}
\begin{figure}
\bc
\includegraphics[width=\columnwidth,clip]{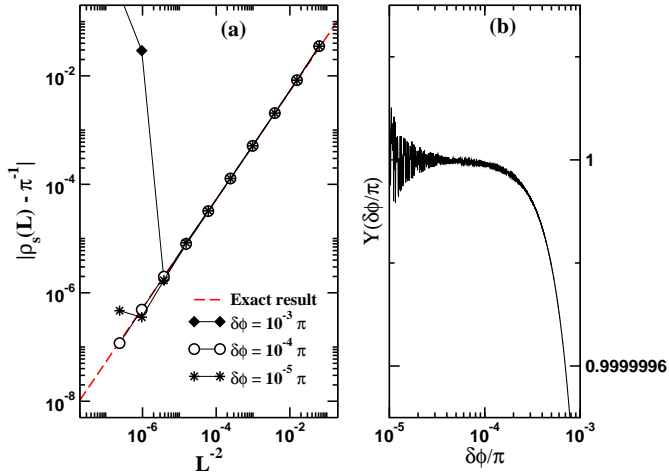}
\caption{(a) Magnitude of the FS corrections of the spin stiffness $|\rho_{S}(L)-\pi^{-1}|$ for different choices of the twist angle $\phi$ calculated for the pure XX model [Eq.(\ref{eq:DefPureXX})]. The long-dashed line is the exact result $L\sin(\frac{\pi}{L})^{-1}-\pi^{-1}$ and the different symbols show the numerical results for different values of the twist. (b) Function $Y(\delta\phi)=2 \frac{1-\cos(\delta \phi)}{(\delta \phi)^2}$ computed in double precision type.}
\label{fig:FSC.Pure}
\ec
\end{figure}
%
Numerical estimates for the spin stiffness can be obtained by 
approximating Eq.(\ref{eq:stiff.def}) for finite L by
\be
\label{eq:stiff.def.disc}
\rho_S\simeq 2 \frac{\epsilon_{0}(\phi)-\epsilon_{0}(0)}{(\delta \phi)^2},
\ee
where $\delta \phi=\phi/L$ is the twist per site. Hence for a given
system the calculation of $\rho_S$ requires to compute
Eq.(\ref{eq:GSEn}) twice: once for finite $\phi$ and once for
$\phi=0$. Since the corrections are of order $1/L^2$ an extrapolation
$L\to\infty$ is, in principle, straightforward and yields the desired
result. However, the appropriate choice of $\phi$ is somewhat delicate
as we show in Fig.\ref{fig:FSC.Pure} (a). Here the numerical results
for the FS scaling of the spin stiffness of the pure chain are
depicted, computed for various system sizes ($L=4,8,16,...,2048$) with
three different values of the twist angle, and compared to the exactly
known result given by Eq.(\ref{FSSS}).  The discrepancy between the
numerical data and the exact result, observed for $\delta\phi/\pi
=10^{-3}$ and $\delta\phi/\pi =10^{-5}$ can be understood as
follows. Using Eq.(\ref{eq:GsPureEnPhi}) one can rewrite
Eq.(\ref{eq:stiff.def.disc}) as
\be
\label{eq:stiff.def.disc2}
\rho_S\simeq 2 \frac{1-\cos(\delta \phi)}{(\delta \phi)^2} J (L \sin{\frac{\pi}{L}})^{-1}.
\ee
The function $Y(\delta\phi)=2 \frac{1-\cos(\delta \phi)}{(\delta
\phi)^2}$, which is exactly equal to one when $\delta\phi=0$, 
is expected to decrease slowly when $\delta\phi$ increases. However,
the numerical calculation of $Y(\delta\phi)$ is limited by the machine
precision and therefore we observe in Fig.\ref{fig:FSC.Pure} (b) that
even in double precision type, for $\delta\phi/\pi <10^{-4}$
undesirable oscillations appear. This puts a bound for the smallest
value of $\delta\phi$ that is meaningful for our numerical procedure,
as we demonstrate in Fig.\ref{fig:FSC.Pure} (a) for
$\delta\phi/\pi=10^{-5}$. On the other hand, when $\delta\phi >
10^{-4}$ the value of $Y$ deviates significantly from one
as shown in Fig.\ref{fig:FSC.Pure} (a) for $\delta\phi/\pi=10^{-3}$. 
Therefore, for a numerical
calculation in double precision, the numerical derivative
Eq.(\ref{eq:stiff.def.disc}) gives the most reliable results for
$\delta\phi \simeq 10^{-4}\pi$ which is confirmed by the numerical
data obtained in this case, shown in Fig.\ref{fig:FSC.Pure} (a).
\section{Localization transition : Scaling from pure to infinite 
randomness behavior}
\subsection{Bosonization predictions for weak disorder and scaling 
argument in the localized-random singlet phase}

The critical behavior of the XXZ model [Eq.(\ref{eq:DefHamXXZ})] with
weak randomness in the couplings and/or in external magnetic fields
has been studied by Doty and Fisher \cite{Doty92} using a bosonization
approach. They found that for random perturbations which preserve the
XY symmetry, the critical properties belong to the universality class
of Giamarchi-Schulz transition for 1D bosons in a random potential
\cite{Giamarchi88}. Let us define the disorder parameter $\mathcal D$
by
\be
\label{eq:DefD}
{\mathcal{D}}={\overline{(J_{i})^{2}}}-\Bigl({\overline{J_{i}}}\Bigr)^2.
\ee
For weak initial randomness $\mathcal D_{0} \ll 1$, the renormalization of the disorder under a change of length scale $l=\ln L$ is \cite{Doty92,Giamarchi88} 
\be
\label{RG}
\frac{\partial {\mathcal{D}}}{\partial{\ln L}}=(3-2K){\mathcal{D}},\\
\ee
where $K$ is the $\Delta$-dependent Luttinger liquid parameter
$K(\Delta)=\frac{\pi}{2(\pi-\mu)}$. Therefore, if $K<3/2$
(i.e. $-\frac{1}{2}<\Delta<1$) the disorder is a relevant perturbation
and the line of pure fixed points is unstable under any amount of
randomness. Under renormalization the system runs into an infinite
randomness fixed point (IRFP) \cite{Doty92,Fisher94}. Using a real
space decimation procedure
\cite{Fisher94}, Fisher reached the same conclusion and demonstrated
analytically the existence of an attractive IRFP.  Strictly speaking,
at the IRFP the system is in the so-called random singlet phase (RSP) or
in the fermionic language, the fermions are localized and their
transport properties are the ones of an insulator. For instance, the
Drude weight is expected to be $0$ in the thermodynamic limit
$L\to\infty$. The renormalization flow is controlled by a disorder
dependent length scale which emerges from Eq.(\ref{RG}), the
localization length:
\be
\label{xi}
\xi^*({\mathcal{D}})\sim {\mathcal{D}}^{-\frac{1}{3-2K}}.
\ee
In the thermodynamic limit the spin stiffness is finite in the QLRO
phase (see Eq.(\ref{eq:stif.pure})) and its FS scaling behavior is
well known \cite{laflo01}. On the other hand, when $\mathcal D >0$ we
have $\rho_S(L,{\mathcal{ D}}) \to 0$ and expect a scaling of the form
\be
\label{FSC}
\rho_S(L,{\mathcal D})=g(\frac{L}{\xi^*(\mathcal D)}),
\ee
with $g$ a universal function. Defining
$x=L/\xi^*(\mathcal D)$, one can consider 3 different regimes : 
(i) For $x \ll 1$, i.e.\ on small length scales, the systems appears to 
    be delocalized with $g \simeq \pi^{-1}$. 
(ii) For $x \gg 1$, i.e. on large length scales, the system shows the 
     asymptotic behavior of the IRFP and $g \to 0$. 
(iii) In the intermediate region $x\sim 1$, a crossover between the 
      pure repulsive fixed point and the attractive IRFP occurs. 
Utilizing standard FS scaling arguments \cite{Wallin94}, one can predict
the behavior of $g(x)$ in the asymptotic regime of the IRFP: $\rho_S$
has dimension of inverse (${\rm length}^{d-2}\times\xi_\tau$), where
$\xi_\tau$ is the correlation length in the imaginary time direction
\cite{Wallin94}. In our case $\xi_\tau\sim\exp(A\xi^{1/2})$, which is
one manifestation of the critical behavior at the IRFP (i.e. the
dynamical exponent formally being $z=\infty$), and $\xi=L$ for a
finite system at criticality. Therefore we expect $\rho_S$ to scale as
\cite{note1}
\be
\label{eq:ScalIRFP}
\ln \rho_S(L) \sim -{\sqrt{L}}.
\ee
Combining this with Eq.(\ref{FSC}), we expect $g(x)$ to behave as a
constant $\simeq \pi^{-1}$ in the delocalized regime (i) and to vanish
as
\be
\label{eq:scalingG}
\ln g(x) \sim -\sqrt{x}
\ee
in the localized regime (ii).

\subsection{Numerical results}
\begin{figure}
\bc
\includegraphics[width=\columnwidth,clip]{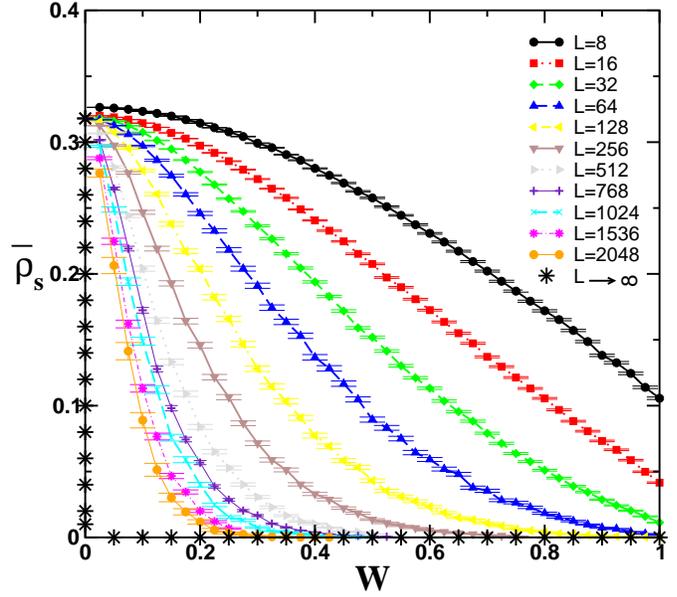}
\caption{Disorder averaged value of the spin stiffness 
${\overline {\rho_S}}$ vs the disorder strength $W\sim \sqrt{{\mathcal
D}}$ for different system sizes, as indicated on the plot. Averaging
has been done over $N_s=10^3$ samples for the biggest sizes and up to
$10^5$ for the smallest ones such that the error bars are well
controlled, as we can observe. The expected behavior in the
thermodynamic limit is represented by the black stars.}
\label{fig:Sti.rand}
\ec
\end{figure}
Following the method explained in Sec.~2., we study the
spin-$\frac{1}{2}$ XX model [Eq.(\ref{eq:DefReXX})] with random bonds
$J_i$ distributed according to the flat distribution
\be
\label{DistDis}
{\mathcal{P}}(J)=\left\{
\begin{array}{rl}
\frac{1}{2W} & \mathrm{if}\ J\in [1-W,1+W]\\
0 & {\mathrm {otherwise.}}\
\end{array}
\right.,
\ee
which implies that the disorder strength is 
${\mathcal D}=\frac{1}{3} W^2$.
Due to the strong sample-to-sample fluctuations that occur in many
disordered quantum systems at low or zero temperatures we have to
perform a disorder average over a sufficiently large number of
samples. In our calculations the latter ranges from $N_s=10^3$ for the
biggest size up to $10^5$ for the smaller ones such that the error
bars are well controlled, as we checked carefully. The system sizes
vary from $L_{min}=8$ to $L_{max}=2048$ and we considered a large
range of disorder strengths between $W_{min}=0.025$ and
$W_{max}=1$. The spin stiffness $\rho_{S}$ was evaluated using
Eq.(\ref{eq:stiff.def.disc}) with a twist angle $\phi=L \times \delta
\phi=L\times \pi/10^4$ and was then averaged over $N_s$ independent
samples:
${\overline{\rho_S}}=\frac{1}{N_s}\sum_{\{\mathrm{samples}\}}\rho_S$. 
\begin{figure}
\bc
\includegraphics[width=\columnwidth,clip]{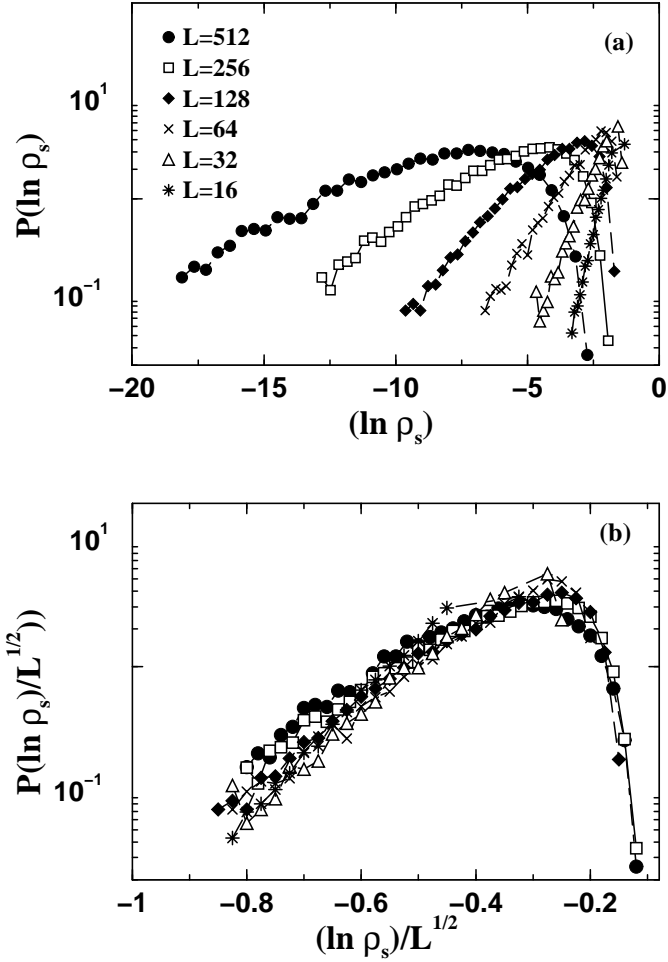}
\caption{Distribution of the spin stiffness obtained with $N_s=10^4$ 
samples at $W=0.5$. The system sizes are indicated on the plot. (a)
The distribution of $\ln(\rho_S)$ is broadening with system size. (b)
Scaling plot of the data shown in Fig.\ref{fig:Dist.Sti} (a), assuming
that the logarithm of the stiffness varies as the square root of the
system size.}
\label{fig:Dist.Sti}
\ec
\end{figure}
In Fig.\ref{fig:Sti.rand} ${\overline{\rho_S(W)}}$ is shown for
different system sizes and we see clearly that it approaches zero for
increasing $L$. In order to validate the FS scaling form
[Eq.(\ref{eq:ScalIRFP})], we studied the distribution of $\ln \rho_S$.
For $W=0.5$, Fig.\ref{fig:Dist.Sti} (a) shows such a distribution for
system sizes ranging from 8 to 512 sites with $N_s =10^4$ samples. As
expected for a system described by an IRFP the distribution gets
broader with increasing system size, which confirms that the dynamical
exponent is formally infinite $z=\infty$. Following
Eq.(\ref{eq:ScalIRFP}), the distribution $P(\frac{\ln
\rho_S}{\sqrt{L}})$ is plotted in Fig.\ref{fig:Dist.Sti} (b) and as
expected, the data of Fig.\ref{fig:Dist.Sti} (a) collapse in a
universal function.
\begin{figure}
\bc
\epsfig{file=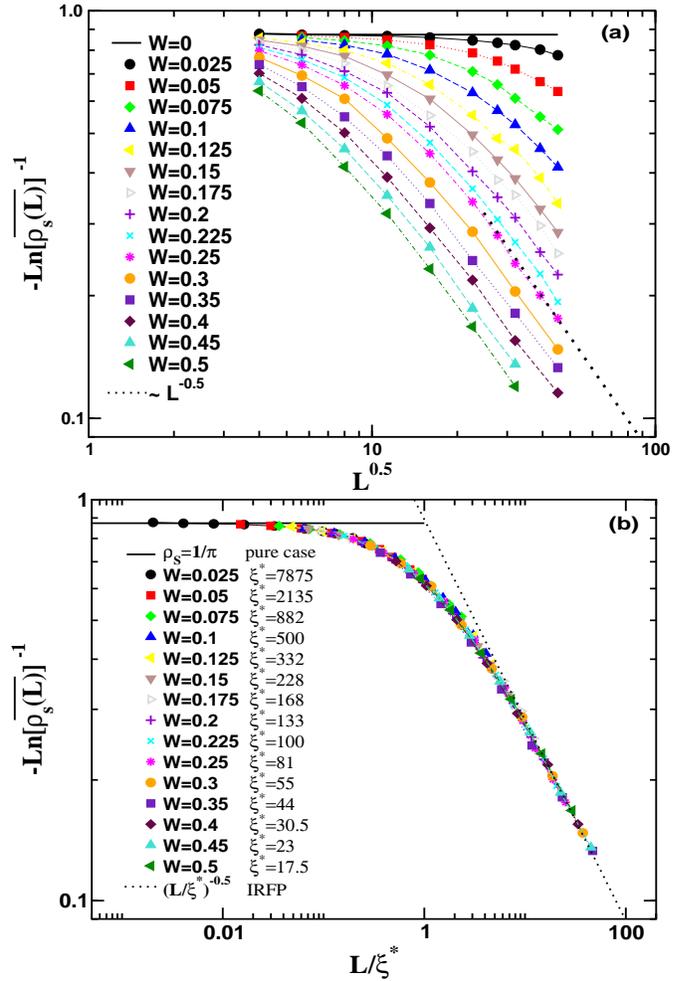,height=13cm,width=\columnwidth,clip} 
\caption{(a) Inverse logarithm of the disorder averaged spin 
stiffness plotted for several box sizes $W$ specified on the plot. The
error bars are smaller than symbol sizes. The full line stands for the
pure case and the dotted one shows the expected IRFP behavior
[Eq.(\ref{eq:ScalIRFP})]. (b) Scaling plot according to Eq.(\ref{FSC})
of the data shown in Fig.\ref{fig:sti.scale}(a) with $\xi^*$ indicated
on the plot for each $W$. Pure and IRFP behavior are indicated
respectively by full and dotted lines.}
\label{fig:sti.scale} 
\ec
\end{figure}

When the disorder is weaker, we expect strong FS effects and a
disorder-dependent length scale might control a crossover 
between the pure repulsive XX fixed point and the attractive
IRFP. Such a behavior is illustrated in Fig.\ref{fig:sti.scale}(a)
since ${\overline{\rho_S(L)}}$ has been calculated for various values
of the disorder $W$. Typically, when $W\ge 0.3$ we can observe the
asymptotic behavior $\ln {\overline{\rho_S(L)}}\sim -L^{1/2}$ as soon
as $L\simeq 100$ but when $W < 0.1$ the pure behavior
${\overline{\rho_S}}\simeq \pi^{-1}$ remains robust up to very large
$L$ and even for $L=2048$ the IRFP asymptotic regime is not yet
reached. 

In order to characterize this crossover behavior, we studied the
scaling function defined by Eq.(\ref{FSC}) and a corresponding scaling
plot of $-(\ln g(L/\xi^*))^{-1}$ is shown in
Fig.\ref{fig:sti.scale}(b). For $W=0.225$ we have chosen $\xi^*=100$
such that the crossover region is centered around $x=L/\xi^*\simeq 1$
and the other estimates, indicated on the plot, have been adjusted
carefully in order to obtain the best data collapse. The 3 regimes
mentioned above (see Sec.~3.1) are clearly visible : The pure regime
(i) for which the stiffness takes values close to $\pi^{-1}$ is
observed if $x \ll 1$.  When $x \gg 1$ the infinite randomness regime
(ii) is relevant : the universality of the IRFP is recovered and $g(x)$
is in perfect agreement with Eq.(\ref{eq:scalingG}). The intermediate crossover 
regime (iii) is visible for $x\sim 1$.

\section{The localization length as a crossover length scale}

\begin{figure}
\bc  
\epsfig{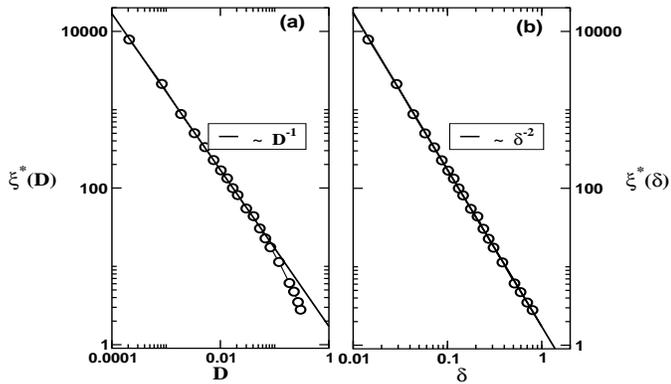} 
\caption{Disorder dependence of the localization length $\xi^*$ of the 
random XX chain. Numerical results are shown with open circles and
full lines represent power-laws as indicated on the plot. (a) As a function of
the disorder parameter ${\mathcal{D}}$ and (b) as a function of the
disorder parameter $\delta$.}
\label{fig:xi.sti} 
\ec
\end{figure}

Finally, we study the disorder dependence of the localization length
$\xi^*$. Using the values extracted from the data collapse shown in
Fig.\ref{fig:sti.scale}(b), $\xi^*({\mathcal D})$ is shown in
Fig.\ref{fig:xi.sti}(a) for several values of the disorder
strength. The numerical results are compared with the predicted
power-law behavior Eq.(\ref{xi}) which is at the XX point given by
$\xi^*({\mathcal D})\sim {\mathcal D}^{-1}$. The agreement between the
numerical results and the bosonization prediction is very good for
weak disorder, but for ${\mathcal D} >0.1$ the data deviate from a
power-law. In order to extract a functional form for $\xi^*$ also in
this range of disorder we look at its behavior as a function of the
variance $\delta$ of the random variable $\ln J_i$:
\be
\label{delta}
\delta={\sqrt{{\overline{(\ln J_i)^{2}}}-\Bigl({\overline{\ln J_i}}\Bigr)^{2}}}
\ee
which is related to $W$ via
\be
\delta={\sqrt{1-\frac{1-W^2}{4W^2}\Bigl[\ln\Bigl(\frac{1+W}{1-W}\Bigr)\Bigr]^2}}.
\label{relwdel}
\ee

As we can observe in Fig.\ref{fig:xi.sti}(b), the parameter $\delta$
is very useful to describe the disorder dependence of $\xi^*$ for any
strength of randomness, indeed the power-law $\xi^*(\delta)\sim
\delta^{-2}$ works perfectly for the whole range of randomness
considered here. Hence we assume that Eq.(\ref{xi}) has to be
replaced, for strong disorder, by
\be
\label{eq:xi.delta}
\xi^{*}(\delta)\sim \delta^{-\Phi},
\ee
and since for weak disorder $\delta\sim \sqrt{\mathcal D}$, we expect
$\Phi=\frac{2}{3-2K}$.

\begin{figure}
\bc
\epsfig{file=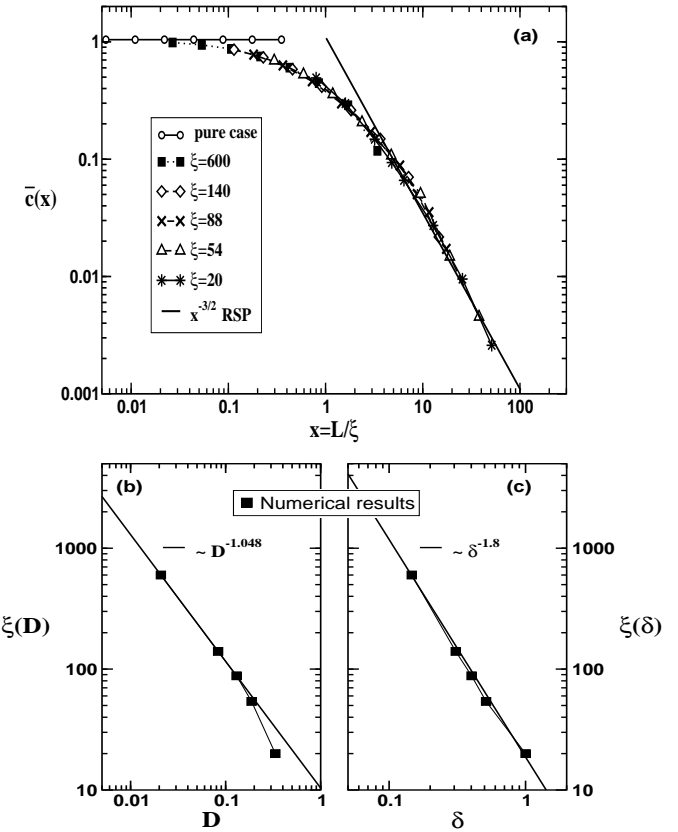,width=\columnwidth,height=11cm,clip} 
\caption{(a) Scaling plot according to Eq.(\ref{eq:corr.scal}) of 
mid-chain $xx$ correlation function data obtained in
\cite{comment,laflo03} for 5 different values of $W$ indicated on the
figure as well as the $\xi$ used for the data collapse. The line with
open circles shows the pure behavior and the full line shows the RSP
behavior at the IRFP. The crossover length scale $\xi$ is plotted vs
$\mathcal D$ and fitted by ${\mathcal D}^{-1.048\pm 0.1}$ only for
weak disorder in (b) whereas in (c) $\xi(\delta)$ displays a better
agreement with a power law $\sim \delta^{-1.8\pm 0.2}$, $\forall
\delta$.}
\label{fig:Corr}
\ec
\end{figure}

Actually, a similar conclusion was drawn in
\cite{comment,laflo03}, where the crossover effects visible in
the spin-spin correlation function of random AF spin chains were
studied. Indeed the correlation functions of the weakly disordered
spin-$\frac{1}{2}$ XXZ chain display a strong crossover behavior
controlled by a disorder-dependent crossover length scale $\xi$ which
behaves as $\delta^{-1.8\pm 0.2}$ \cite{comment}.  In analogy to what
we did with the stiffness above, we can extract the crossover length
scale $\xi$ using the scaling function
\be
\label{eq:corr.scal}
{\tilde c}(x)={\mathcal C}_{\mathrm avg}(L)/{\mathcal C}_{0}(L),~~{\mathrm{with}}~x=\frac{L}{\xi},
\ee
where ${\mathcal C}_{0}(L)$ and ${\mathcal C}_{\mathrm avg}(L)$ are
spin-spin correlation functions calculated at mid-chain respectively
for the pure and random models. At the XX point, when $W=0$,
${\mathcal C}_{0}(L)\propto L^{-1/2}$ and at the IRFP ${\mathcal
C}_{\mathrm avg}(L)\propto L^{-2}$. The crossover between these two
distinct behaviors is shown in Fig.\ref{fig:Corr}(a) where ${\tilde
c}(x)$ presents a universal form, following ${\tilde c}(x)=$constant
for $x\ll1$ and ${\tilde c}(x)\sim x^{-3/2}$ for $x\gg 1$.  We see
that the characteristic length scale $\xi$ beyond which the asymptotic
IRFP behavior sets in in the correlation function scales with 
disorder strength in very much the same way as the localization length
$\xi^*$.

\section{Conclusion}

In this paper we have studied the scaling behavior of the stiffness of
the random antiferromagnetic spin-$\frac{1}{2}$ XX chain numerically
via exact diagonalization calculations utilizing the fact that the
system can be mapped on a free fermion model. The latter allowed us to
study rather large system sizes by which we were able to analyze
thoroughly the crossover effects observable for weak disorder.  Our
results clearly show that the asymptotic behavior of the model under
consideration is governed by an infinite randomness fixed point for
all disorder strengths, including the weakest, as predicted by
D. Fisher \cite{Fisher94}. We could observe one of the characteristics
of the IRFP, namely a formally infinite value for the dynamical
exponent, from the finite size scaling behavior of the probability
distribution of the stiffness, where $\ln\rho_s/L^{1/2}$ occurs as a
scaling variable indicating that the stiffness scales exponentially
with the the square root of the system size. 

Moreover we showed that the finite size scaling form of the average
value of the stiffness is governed by a characteristic length scale
that depends on the strength of the disorder. The length scale can be
identified as a localization length with regard to transport
properties but also as a crossover length scale below which the system
behaves essentially like a pure (disorder free) chain and the
stiffness is constant and beyond which the asymptotic behavior
characteristic for an infinite randomness fixed point becomes
visible and the stiffness scales to zero with a characteristic power
of the system size. We found that this length scale diverges like
$1/\delta^2$ with decreasing variance $\delta$ of the disorder, which
agrees well with an analytical prediction using bosonization
techniques. This behavior agrees also well with the scaling behavior
of the crossover length for the spin-spin correlation function, which
indicates that there is indeed a single disorder strength dependent
length scale governing the crossover as well as the localization
phenomena in this system.


\begin{thebibliography}{9}
\bibitem{noteNum} The linear algebra package LAPACK has been used here.
\bibitem{note1} Strictly speaking there is a logaritmic correction but it is subdominant.
\end{thebibliography}
\end{document}